\documentclass[aps,prd,nofootinbib,preprint,floatfix,unsortedaddress]{revtex4-1}
\usepackage{setspace}
\usepackage{graphicx}
\usepackage{dcolumn}
\usepackage{multirow}
\usepackage{subfigure}
\usepackage{times,mathptm}
\usepackage{float}
\usepackage{color}
\usepackage{amsmath,amsfonts}
\usepackage{mathptmx}
\usepackage{mathrsfs}
\usepackage{bbm}
\usepackage{bm}
\usepackage{xfrac}

\newcommand{\bea}{\begin{eqnarray}}
\newcommand{\eea}{\end{eqnarray}}

\renewcommand{\d}{\delta}

\newcommand{\ihat}{\boldsymbol{\hat{\textbf{\i}}}}

\renewcommand{\l}{\lambda}

\newcommand{\kL}{k_L}

\renewcommand{\b}{\beta}

\newcommand{\tr}{\text{Tr}}

\newcommand{\bx}{\mathbf{x}}

\newcommand{\vx}{{\vec{x}}}
\newcommand{\vy}{{\vec{y}}}

\newcommand{\vk}{{\vec{k}}}

\newcommand{\m}{\mu}
\newcommand{\g}{\gamma}

\newcommand{\s}{\sigma}
\renewcommand{\k}{\kappa}
\newcommand{\V}{{\cal V}}
\newcommand{\D}{\Delta}

\newcommand{\Z}{{\cal Z}}

\renewcommand{\th}{\theta}

\newcommand{\oh}{\frac{1}{2}}
\newcommand{\oq}{\frac{1}{4}}

\newcommand{\dg}{\dagger}
\newcommand{\non}{\nonumber}

\newcommand{\rf}[1]{(\ref{#1})}
\newcommand{\ra}{\rightarrow}

\renewcommand{\vec}[1]{\bm #1}

\usepackage{ulem}

\begin{document}

\bibliographystyle{h-physrev5}

\title{The potential of the effective Polyakov line action from the underlying lattice gauge theory} 
 
\author{Jeff Greensite}
\affiliation{Niels Bohr International Academy, Blegdamsvej 17, DK-2100
Copenhagen \O, Denmark}
\altaffiliation[Permanent address: ]{Physics and Astronomy Dept., San Francisco State
University, San Francisco, CA~94132, USA}
\date{\today}
\vspace{60pt}
\begin{abstract}

\singlespacing
 
    I adapt a numerical method, previously applied to investigate the Yang-Mills vacuum wavefunctional, to the problem
of extracting the effective Polyakov line action from SU($N$) lattice gauge theories, with or without matter fields.  The method can be used to find the variation of the effective Polyakov line action along any trajectory in field configuration space; this information is sufficient to determine the potential term in the action, and strongly constrains the possible form of the kinetic term. The technique is illustrated for both pure and gauge-Higgs SU(2) lattice gauge theory at finite temperature. A surprise, in the pure gauge theory, is that the potential of the corresponding Polyakov line action contains a non-analytic (yet center-symmetric) term proportional to $|P|^3$, where $P$ is the trace of the Polyakov line at a given point, in addition to the expected analytic terms proportional to even powers of $P$.   
\end{abstract}

\pacs{11.15.Ha, 12.38.Aw}
\keywords{Confinement,lattice
  gauge theories}
\maketitle

\singlespacing
\section{\label{sec:intro}Introduction}

     Consider a lattice gauge theory with gauge group SU($N$) on a periodic lattice of time extent $N_t$, possibly containing matter fields and a chemical potential.  If we integrate out all degrees of freedom under the constraint that Polyakov line holonomies are held fixed, then the resulting distribution depends only on those Polyakov line holonomies or, more precisely, on their eigenvalues.  The logarithm of this distribution is defined to be the effective Polyakov line action 
$S_P$.  

    If the underlying lattice gauge theory in $D=4$ dimensions has a sign problem due to a non-zero chemical potential, then $S_P$ probably also has a sign problem.  However, there are indications that the sign problem may be more tractable in $S_P$ than in the underlying theory. Using strong-coupling and hopping parameter expansions, it is possible to actually carry out the integrations over gauge and matter fields mentioned above, to arrive at an action of the form~\footnote{This is the action at leading order.  For the effective action determined at higher orders in the combined strong-coupling and hopping parameter expansions, cf.\  \cite{Fromm:2011qi}.}
\bea
S_P &=&   \b_P \sum_{\vx} \sum_{i=1}^3  [\tr U_\vx^\dg \tr U_{\vx+\ihat} + \tr U_\vx  \tr U^\dg_{\vx+\ihat}] 
  + \k \sum_\vx [e^\m \tr U_\vx + e^{-\m} \tr U^\dg_\vx]  \ ,
\label{action1}
\eea   
where $\b_P, \k$ are calculable constants depending on the gauge coupling, quark masses, and temperature $T=1/N_t$ in the underlying theory.  To minimize minus signs later on, the overall sign of $S_P$ is defined such that the Boltzmann weight is proportional to $\exp[S_P]$, rather than $\exp[-S_P]$.  The Polyakov line holonomies $U_\vx \in $ SU($N$) in 
\rf{action1} are also known as ``effective spins." A path integral based on an effective spin action of the form \rf{action1}, for a wide range of $\b_P,\k,\m$, can be treated by a number of different methods, including the ``flux representation"  \cite{Mercado:2012ue}, reweighting 
\cite{Fromm:2011qi}, and stochastic quantization \cite{Aarts:2011zn}.  Even traditional mean field methods have had some degree of success in determining the phase diagram  \cite{Greensite:2012xv}.
   
    The problem, of course, is that strong lattice coupling and heavy quark masses lie outside the parameter range of phenomenological interest, and it is not obvious how to extract $S_P$ for parameters inside the range of interest, even at $\mu=0$.  There have been some efforts in this direction, notably the inverse Monte Carlo method of ref.\  
\cite{Wozar:2007tz,*Heinzl:2005xv}, as well as early studies \cite{Gocksch:1984ih,Ogilvie:1983ss} which employed microcanonical and Migdal-Kadanoff methods, respectively.  There is also a strategy for determining the phase structure of lattice gauge theory from an effective spin theory, whose form is suggested by high-order strong-coupling and hopping parameter expansions \cite{Fromm:2011qi}.    Here, however, I will discuss a different approach to the problem, recently suggested in ref.\  \cite{Greensite:2012xv},  which will be illustrated for SU(2) pure gauge and gauge-Higgs theories.

\section{\label{sec:method}The ``Relative Weights" Approach}   

     Let $S_{QCD}$ be the lattice QCD action at temperature $T=1/N_t$ in lattice units, with lattice gauge coupling $\b$, and a set of quark masses denoted collectively $m_q$.  We set chemical potential $\m=0$ for now.  It is convenient to impose a temporal gauge condition in which the timelike link variables are set to the unit matrix everywhere except on a single time slice, say at $t=0$.  In that case, $U_0(\vx,0)$ is the Polyakov line holonomy passing through the site $(\vx,t=0)$.  The effective Polyakov line action
is defined in terms of the partition function
\bea
Z(\b,T,m_q) &=& \int DU_0(\vx,0) \int DU_k D\overline{\psi} D\psi ~ e^{S_{QCD}}
\non \\
                   &=& \int DU_0(\vx,0) ~ e^{S_P[U_0]} \ ,
\label{S0}
\eea
or equivalently
\bea
\exp\Bigl[S_P[U_{\vx}]\Bigl] =    \int  DU_0(\vx,0) DU_k D\overline{\psi} D\psi ~ \left\{\prod_{\vx} \d[U_{\vx}-U_0(\vx,0)]  \right\}
 e^{S_{QCD}} \ .
\label{S_P}
\eea
Because temporal gauge has a residual symmetry under time-independent gauge transformations, it follows that
$S_P[U_{\vx}]$ is invariant under $U_{\vx} \ra g(\vx) U_{\vx} g^\dg(\vx)$, which means that $S_P$ only depends on the
eigenvalues of the Polyakov line holonomies.  

     Now consider a finite set of $M$ SU($N$) ``effective spin"  configurations in the three-dimensional cubic lattice $V_3$ of volume $L^3$, 
 \bea
 \Bigl\{ \{U^{(i)}_{\vx}, \mbox{all~} \vx \in V_3\}, ~ i=1,2,...,M \Bigr\} \ .
\eea 
Each member of the set can be used to specify the timelike links on the timeslice $t=0$.  Define
\bea
\Z =  \int DU_0(\vx,0) DU_k D\overline{\psi} D\psi ~ \sum_{i=1}^M \left\{ \prod_{\vx} \d[U^{(i)}_{\vx} - U_0(\vx,0)]  \right\}
 e^{S_{QCD}} \ ,
\eea
and consider the ratio
\bea
 { \exp\Bigl[S_P[U^{(j)}]\Bigr] \over  \exp\Bigl[S_P[U^{(k)}]\Bigr] }
    &=& { \int DU_0(\vx,0) DU_k D\overline{\psi} D\psi ~ \left\{\prod_{\vx} \d[U^{(j)}_{\vx} - U_0(\vx,0)]  \right\}
 e^{S_{QCD}}  \over
 \int DU_0(\vx,0) DU_k D\overline{\psi} D\psi ~ \left\{ \prod_{\vx} \d[U^{(k)}_{\vx} - U_0(\vx,0)]  \right\}
 e^{S_{QCD}} }
 \non \\
 &=& {  {1\over \Z} \int DU_0(\vx,0) DU_k D\overline{\psi} D\psi ~ \left\{\prod_{\vx} \d[U^{(j)}_{\vx}-U_0(\vx,0)]  \right\}
 e^{S_{QCD}}  \over {1\over \Z} 
 \int DU_0(\vx,0) DU_k D\overline{\psi} D\psi ~ \left\{\prod_{\vx} \d[U^{(k)}_{\vx}-U_0(\vx,0)]  \right\}
 e^{S_{QCD}} }  \ ,
 \eea
 where in the second line we have merely divided both the numerator and denominator by a common factor.
 However, by inserting this factor, both the numerator and denominator acquire a meaning in statistical mechanics, because the factor  $\Z$ can be interpreted as the partition function of a system in which the configuration of timelike link variables at $t=0$ is restricted to belong to the set $\{U^{(i)},~i=1,...,M\}$.  This means that 
 \bea
 \mbox{Prob}[U^{(j)}] =  {1\over \Z} \int DU_0(\vx,0) DU_k D\overline{\psi} D\psi ~ \left\{\prod_{\vx} 
 \d[U^{(j)}_\vx-U_0(\vx,0)]  \right\} e^{S_{QCD}}
 \eea
 is simply the probability, in this statistical system, for the $j$-th configuration $U_0(\vx,0) = U^{(j)}(\vx)$ to be found on the $t=0$ timeslice.  This probability can be determined from a slightly modified Monte Carlo simulation of the original lattice action.  The simulation proceeds by standard algorithms, for all degrees of freedom other than the timelike links
at $t=0$, which are held fixed.  Periodically, on the $t=0$ timeslice, one member of the given set of timelike link configurations is selected by the Metropolis algorithm, and all timelike links on that timeslice are updated simultaneously. Let $N_i$ be the number of times that the $i$-th configuration is selected by the algorithm, and ${N_{tot} = \sum_i N_i}$.  Then $\mbox{Prob}[U^{(j)}]$ is given by
\bea
         \mbox{Prob}[U^{(j)}] = \lim_{N_{tot}\ra\infty} {N_j \over N_{tot}} \ ,        
\eea
and this in turn gives us the {\it relative weights} 
\bea
 { \exp\Bigl[S_P[U^{(j)}]\Bigr] \over  \exp\Bigl[S_P[U^{(k)}]\Bigr] } = \lim_{N_{tot}\ra\infty} {N_j \over N_{k}}   
\label{rw}
\eea
for all elements of the set.   A computation of this kind allows us to test any specific proposal for $S_P$, which may be motivated by some theoretical considerations.  But it might also be possible, given data on the relative weights of a variety of different sets, to guess the action that would lead to these results.  In this article we will consider sets of spatially constant Polyakov line configurations, and small plane wave perturbations around a constant background.  This is already sufficient to determine the potential term in $S_P$, and to suggest the form of the full action.

    The method described above was proposed long ago \cite{Greensite:1988rr} in connection with the Yang-Mills vacuum wavefunctional. Recently there have been some sophisticated suggestions for the form of this wavefunctional in 2+1 dimensions, and the technique was revived in order to test these ideas in ref.\ \cite{Greensite:2011pj}. The main difference between the method as applied to vacuum wavefunctionals, and as applied to determining $S_P$, is that in the former case the simulation chooses from a fixed set of spacelike link configurations on the $t=0$ timeslice, while in the latter the choice is made from a set of timelike link configurations.

\subsection{Finite chemical potential}

   Let $S^{\m}_{QCD}$ denote the QCD action with a chemical potential, which can be obtained from $S_{QCD}$ by the following replacement of timelike links at $t=0$:
\bea
S^{\m}_{QCD} = S_{QCD}\Bigr[U_0(\bx,0) \ra e^{N_t \m} U_0(\bx,0), U^\dg_0(\bx,0) \ra e^{-N_t \m} 
U^\dg_0(\bx,0)\Bigl] \ .
\eea    
The corresponding Polyakov line action $S_P^\m$ is in principle obtained from \rf{S_P}, with $S^{\m}_{QCD}$ as the underlying action.  Of course the integration indicated in \rf{S_P} can so far only be carried out for strong couplings and large quark masses,  but it is not hard to see that each contribution to $S_P$ in the strong-coupling + hopping parameter expansion at $\m=0$ maps into a corresponding contribution to $S^\m_P$ by the replacement
\bea
U_\vx \ra e^{N_t \m} U_\vx ~~~,~~~ U^\dg_\vx \ra e^{-N_t \m} U^\dg_\vx \ .
\label{replace}
\eea
It is reasonable then to suppose that this mapping holds in general, i.e.\ if we have by some means obtained 
$S_P[U_\vx,U^\dg_\vx]$ beyond the range of validity of the strong-coupling + hopping parameter expansion, then
the corresponding $S^\m_P$ is obtained by making the change of variables \rf{replace}.  There is, however, a
possible source of ambiguity in this scheme (noted in \cite{Greensite:2012xv}), coming from identities such as
\bea
\tr U^\dg_\vx = \oh \Bigl[ (\tr U_\vx)^2 - \tr U_\vx^2 \Bigr]
\label{identity}
\eea
in SU(3).  One way around this ambiguity is to enlarge the range of $U_0(\vx,0)$, allowing these variables to take
on values 
\bea
         U_0(\vx,0) = e^{i\th} U(\vx) \ ,
\eea
where $U(\vx)$ is an element of $SU($N$)$.  In other words, we allow the $U_0(\vx,0)$ links to take on values
in the $U(N)$ group, although it will be sufficient for our purposes to let $\theta$ be 
$\vx$-independent.\footnote{It is also sufficient to restrict $\th$ to $0\le \th < 2\pi/N$.  The full range $[0,2\pi]$
is redundant, because of the $Z_N$ center of SU($N$).}  Suppose we
are able to determine $S_P$ for this enlarged domain of Polyakov line variables.  Then $S_P^\m$ is obtained
by analytic continuation, $\th \ra -iN_t \m$.

    The essential point here is that if one can determine $S_P$ by simulations of $S_{QCD}$ at $\m=0$, then this result can be used to determine $S_P^\m$ at finite chemical potential.  If the sign problem is in fact tractable for $S_P^\m$, 
as recent results seem to suggest, then this may be a useful way of attacking the sign problem in full QCD.

\subsection{Relative weights, and path-derivatives of $\mathbf  S_P$}

   Let ${\cal C}$ be the configuration space of effective spins $\{U_\vx\}$ on an $L^3$ lattice, and let the variable $\l$
parametrize some path $\{U_\vx(\l)\}$  through ${\cal C}$.  The method of relative weights is particularly useful in computing derivatives of the Polyakov line action
\bea
            \left({d S_P \over d \l} \right)_{\l = \l_0}
\eea
along the path.   To see this, we begin by taking the logarithm of both sides of eq.\ \rf{rw}, and find 
\bea
         S_P[U^{(j)}] - S_P[U^{(k)}]   &=& \lim_{N_{tot}\ra\infty} \Bigl\{ \log N_j - \log N_k \Bigr\}
\non \\
&=& \lim_{N_{tot}\ra\infty} \left\{ \log {N_j\over N_{tot}} - \log {N_k\over N_{tot}}  \right\} \ .
\eea
(From this point on we will drop the limit.)  Now imagine parametrizing the effective spins by a parameter $\l$; each value of $\l$ gives us a different configuration $U_\vx(\l)$.  Let the configuration $U^{(j)}$ correspond to $\l=\l_0+\D \l$, and 
$U^{(k)}$ correspond to $\l=\l_0-\D \l$.  Then
\bea
            \left({d S_P[U_\vx(\l)] \over d\l}\right)_{\l=\l_0} \approx {1\over 2\D \l}
            \left\{ \log {N_j\over N_{tot}} - \log {N_k\over N_{tot}} \right) \ .
\eea
However, rather than using only two configurations to compute the derivative, we can obtain a more accurate numerical estimate if we let $\l$ increase in increments of $\D \l$, e.g. 
\bea
           \l_n = \l_0 + \left( n - {M+1 \over 2}\right) \D \l ~~~,~~~ n=1,2,...,M ~~~  \ ,
\eea
and use all of the $M$ values obtained for $N_n$ in the simulation. For $\D \l$ small enough, the data for $\log N_n/N_{tot}$ vs.\ $\l_n$ will fit a straight line, and then we obtain the estimate
\bea
\left({d S_P[U_\vx(\l)] \over d\l}\right)_{\l=\l_0} \approx ~~ \mbox{slope of } \log {N_n \over N_{tot}} 
    ~~ \mbox{vs.}~~ \l_n \ .
\eea
The procedure will be illustrated explicitly in the next section.

\section{\label{sec:test}Testing the method at strong coupling} 
    
    The first step is to compute $d S_P/d \l$ for a case where we know the answer analytically.  As mentioned previously, $S_P$ can be readily computed in the strong-coupling + hopping parameter expansion.  We will consider here the case of pure SU(2) Yang-Mills theory at a strong coupling $\b$.  If the lattice is $N_t$ lattice spacings in the time direction,
then computing the diagrammatic contributions to $S_P$ at leading and next-to-leading order in the 
strong-coupling/character expansion we find
\bea
           S_P &=& \left[1 + 4N_t \left({I_2(\b) \over I_1(\b)}\right)^4 \right] 
                          \left({I_2(\b) \over I_1(\b)}\right)^{N_t} \sum_\vx \sum_{i=1}^3 \tr U_\vx \tr U_{\vx + \ihat}
\non \\
&=& \b_P   \sum_\vx \sum_{i=1}^3   P_\vx P_{\vx+ \ihat}  \ ,                
\label{Sp_strong}
\eea                          
where 
\bea
    P_\vx &\equiv& \oh \tr U_\vx
\non \\
    \b_P &=&  4 \left[1 + 4N_t \left({I_2(\b) \over I_1(\b)}\right)^4 \right]  \left({I_2(\b) \over I_1(\b)}\right)^{N_t} \ .
\eea

    Let us first consider sets of spatially constant configurations with varying amplitudes in the neighborhood of 
$P=P_0$, i.e.
\bea
            U^{(n)}_\vx &=& (P_0 + a_n) \mathbbm{1} + i \sqrt{1 - (P_0+a_n)^2} \s_3
\non \\ 
               a_n &=& \Bigl(n - \oh(M+1)\Bigr) \D a ~~,~~ n=1,2,...,M ~~~~ \ ,
\label{path1}
\eea
so in this case $a$ is the $\l$ parameter of the previous section. 
If we divide $S_P$ into a kinetic and potential part, which in the case of \rf{Sp_strong} is
\bea
S_P &=&  K_P + V_P
\non \\
K_P &=& \oh \b_P \sum_\vx \sum_{i=1}^3   (P_\vx P_{\vx + \ihat} - 2P_\vx^2 + P_\vx P_{\vx - \ihat})
\non \\
V_P &=& 3 \b_P \sum_\vx P_\vx^2 \ ,
\label{strong}
\eea
then $dS_P/da = dV_P/dP_0$ is giving us the derivative of the potential piece, which can then be reconstructed, up to an irrelevant constant, by integration.  So the procedure for determining $V_P$ (assuming it were not already known from the strong-coupling expansion) is to compute $dV_P/dP_0$ numerically, fit the results to some appropriate polynomial in $P_0$, and then integrate the fit.

\begin{figure}[t!]
\centerline{\scalebox{0.9}{\includegraphics{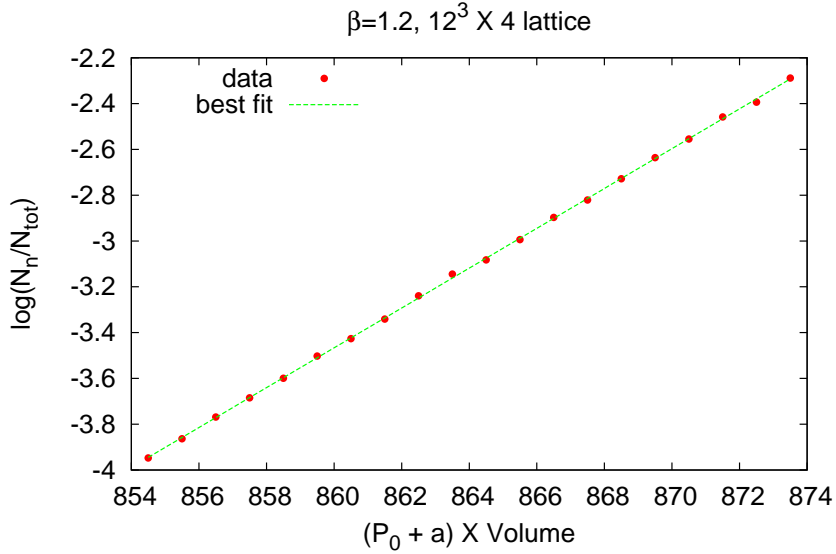}}}
\caption{The slope of the straight-line fit to the data shown gives an estimate for the derivative $L^{-3}dS_P/da$ of $S_P$ with respect to the amplitude of spatially constant effective spin configurations.  In this case, the derivative is evaluated at $P_0=0.5$, for an underlying pure Yang-Mills theory at  strong coupling value of $\b=1.2$, on a $12^3 \times 4$ lattice.}
\label{fig1}
\end{figure}

\begin{figure}[t!]
\centerline{\scalebox{0.9}{\includegraphics{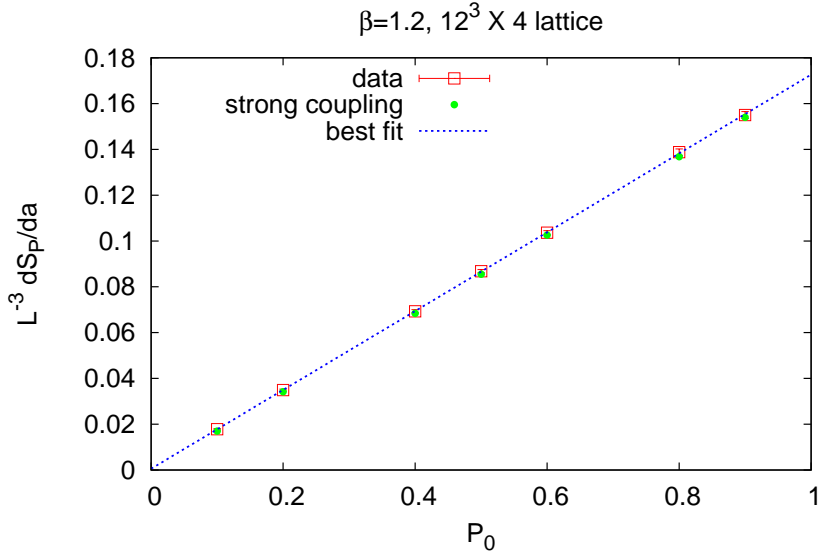}}}
\caption{A plot of the values for $L^{-3} dS_P/da$ vs.\ $P_0$.   Each data point is extracted from a plot similar to the
previous figure.  Also shown are the corresponding strong-coupling values, and a best linear fit to the data points.}  
\label{fig2}
\end{figure}

      Our sample simulation is carried out in pure SU(2) lattice gauge theory at coupling $\b=1.2$ (well within the regime
of strong couplings) on a $12^3 \times 4$ lattice with $M=20$ sets of spatially constant configurations.
Figure \ref{fig1} shows the data for $\log(N_n/N_{tot})$ plotted vs.\ $(P_0+a_n) \times$ spatial lattice volume ($12^3$), at
$P_0=0.5$.  It is clear that the data falls quite accurately on a straight line, and the slope gives an estimate for the derivative
\bea
 {1\over L^3}\left({d S_P(U_\vx(a)) \over da}\right)_{a=0} &=& {1\over L^3} {dV_P(P_0) \over dP_0} 
\label{dSda}
\eea
which can be compared to the value $6 \b_P P_0$ obtained from the strong-coupling expansion.  The derivative obtained from numerical simulation vs.\ $P_0$ is plotted in Fig.\ \ref{fig2}, and it obviously fits a straight line.  Therefore the potential $V_P$ is quadratic in $P_\vx$, and we find, at $\b=1.2$ 
\bea
          V_P = \left\{ \begin{array}{cl}
               0.1721(8) \sum_\vx \oh P_\vx^2  & \mbox{relative weights method} \cr 
                    &  \cr
               0.1710 \sum_\vx \oh P_\vx^2 & \mbox{strong-coupling expansion} \end{array} \right. \ ,
\eea
where we have dropped, in the upper line, an irrelevant constant of integration.  The small numerical difference between the relative weights and strong-coupling results can probably be attributed to neglected higher order terms in the strong-coupling expansion.\footnote{Statistical errors are estimated from best fit slopes obtained from eight independent runs.  Where errorbars are not shown explicitly, in the two-dimensional plots shown below, they are smaller than the symbol size.}
 
    In order to investigate the kinetic term, we consider plane-wave deformations of spatially constant configurations.  The path through configuration space ${\cal C}$ is again parametrized by $a$, with
\bea
            U^{(n)}_\vx &=& P^{(n)}_\vx \mathbbm{1} + i \sqrt{1 -  (P^{(n)}_\vx)^2} \s_3
\non \\ 
               P^{(n)}_\vx &=&  P_0 + a_n \cos(\vk \cdot \vx)
\non \\
                   k_i &=& {2 \pi \over L} m_i \ ,
\label{trajectory}
\eea
where the $\{m_i,~ i=1,2,3\}$ are integers, not all of which are zero.  For this class of configurations we have, for the action
\rf{Sp_strong}
\bea
          S_P = \b_P L^3 \left( 3P_0^2 + \oh a_n^2 \sum_{i=1}^3 \cos(k_i) \right) \ .
\eea
Since the deformation of the action is proportional to $a^2$, it is natural to consider the derivative of $S_P$ with respect
to $a^2$, i.e.
\bea
    {1\over L^3}{dS_P \over d(a^2)} = \oh \b_P \sum_i \cos(k_i)   \ ,
\eea
and therefore we can choose to let $a_n^2$, rather than $a_n$, increase in equal increments, so that 
${a_n = \sqrt{n} \D a}$.
 
\begin{figure}[t!]
\centerline{\scalebox{0.9}{\includegraphics{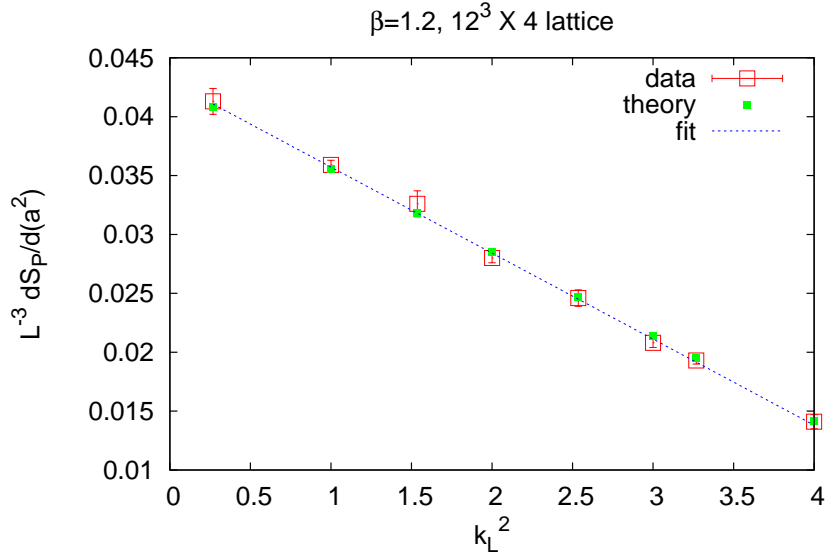}}}
\caption{Derivative of the action w.r.t.\ path parameter $a^2$ vs.\ squared lattice momentum.  Data is taken at
strong gauge coupling $\b=1.2$ for plane-wave deformations. Squares indicate the relative-weights values, while
green dots are the values obtained from the strong-coupling expansion.}  
\label{fig3}
\end{figure}

   The numerical procedure is similar to the determination of the potential term:  we compute the derivative 
$L^{-3}dS_P/d(a^2)$, at fixed $P_0$ and $\vk$, from the slope of a plot of $\log(N_n/N_{tot})$ vs.\ $a_n^2 L^3$.  
Then these values for the derivative are plotted, at various values of $P_0$, against squared lattice momentum
\bea
           \kL^2 \equiv  4 \sum_{i=1}^3 \sin^2(\oh k_i) \ .
\label{k2}
\eea
The result, at $P_0=0.5$, is shown in Fig.\ \ref{fig3}, and we find, for a trajectory \rf{trajectory} at fixed $\vk$,
\bea
           {1\over L^3} {d S_P \over d(a^2)} = -A \kL^2 + B \ ,
\label{deriv}
\eea
where
\bea
A = 7.3(2) \times 10^{-3} ~~,~~ B = 4.30(3) \times 10^{-2} \ .
\eea
The simulation has also been carried out at other values of $P_0$, but the results are almost indistinguishable from Fig.\ \ref{fig3}, and so are not displayed here.  The important point, however, is that the path derivative \rf{deriv} is $P_0$
independent.

    Integrating with respect to $a^2$, we find that along any path parametrized by $a$ with fixed $P_0$
\bea
           S_p[U_\vx(a)] &=& L^3\{ -A a^2 \kL^2  + B a^2 + f(P_0) \}  \ ,
\label{action2}
\eea
where $f(P_0)$ is a constant of integration, which can be determined from the data on the potential:
\bea
         f(P_0) = C P_0^2  ~~,~~  C = 0.0861 \pm 0.0004 \ .
\eea
The next step is to express $S_P$ along the path in terms of $U_\vx$ (or $P_\vx=\oh \tr U_\vx$).   From the definitions 
\rf{trajectory}, \rf{k2}, one easily finds that \rf{action2} can be expressed as
\bea
S_P = 4A \sum_\vx \sum_{i=1}^3 P_\vx P_{\vx+ \ihat} + \Bigl[(B-6A)a^2 + (C-12A) P_0^2 \Bigr] L^3 \ .
\eea
The constants $B-6A$ and $C-12A$ are, within statistical error, consistent with zero.  So we will just drop these terms.  Then along the trajectory the action has the form
\bea
     S_P = (.0292 \pm .0008) \sum_\vx \sum_{i=1}^3 P_\vx P_{\vx+ \ihat}  ~~~\mbox{(relative weights method)}\ ,
\eea     
and of course the natural conjecture is that this is the action itself, at any point in configuration space.  Further checks would be to calculate numerical derivatives $dS_P/d\l$ along other trajectories, to test the consistency of this conjecture.  We don't really need to do that here, since the action at strong couplings is already known analytically, and is given in eq.\ \rf{Sp_strong} to 
leading and next-to-leading order in the strong-coupling expansion.  At $\b=1.2$ we have, from eq.\ \rf{Sp_strong}, that
\bea
    S_P = .0285  \sum_\vx \sum_{i=1}^3 P_\vx P_{\vx+ \ihat}  ~~~\mbox{(strong-coupling expansion)} \ ,
\eea
which is a close match to what we have arrived at via the relative weights procedure.

    This is, perhaps, a lot of effort to derive a known result.  We have gone through this exercise in order to illustrate the method, and to make sure, in a case where the answer is known, that the method actually works.

\section{\label{sec:potential}Potential $\mathbf V_P$ in pure-gauge theory, weaker couplings}     
   
    We now reduce the lattice coupling of the underlying SU(2) pure-gauge theory, setting $\b=2.2$ with inverse temperature $N_t=4$ in lattice units.   At this coupling and temperature (which is still inside the confinement phase of the theory), the effective Polyakov line action $S_P$ is not known.

   The easiest task is to determine the potential part of the action.  For the purposes of this article, we define the kinetic part of the action to be the piece which vanishes for spatially constant configurations, while the potential part is local.
With these definitions
\bea
           V_P &=& \sum_\vx \V(U_\bx) \ ,
\eea
and the function $\V(U_\bx)$ is determined by evaluating $S_P$ on configurations $U_\vx=U$ which are constant in
3-space, i.e.           
\bea
           \V(U) &=& {1\over L^3} S_P(U) \ .
\eea
Then by definition the kinetic part of the action is
\bea
            K_P &\equiv& S_P[U_\vx] - V_P[U_\vx]  \ .
\eea

    In order to determine $V_P$, we consider as before the path through configuration space \rf{path1} parametrized by
the variable $a$, and once again we can identify $dS_P/da$ with $dV_P(P_0)/dP_0$ as in  \rf{dSda}.   The derivatives
are determined by the relative weight method described above, the dependence on $P_0$ is fit to a polynomial, and $V_P$ is then determined, up to an irrelevant constant, by integration over $P_0$. 

    Because the $Z_2$ center symmetry is unbroken at $\b=2.2$ and $N_t=4$, and $\V(U_\vx)$ is a class function, it is natural to assume that $\V(U)$ is well represented by a few group characters $\chi_j(U)$ of zero N-ality ($j=$ integer for SU(2)), and the potential is analytic in $P_\vx$.  Surprisingly, this is {\it not} what is found.  

\begin{figure}[ht]
\centering
\subfigure[~ ${1\over L^3}{dV_P \over dP}$ vs.\ $P$ with a $P^2$ term in the fitting function.]{
\resizebox{90mm}{!}{\includegraphics{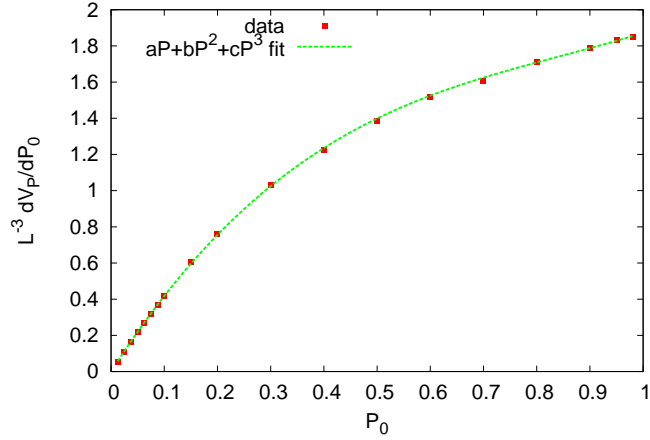}}
\label{sfig4a}
}
\subfigure[~ ${1\over L^3}{dV_P \over dP}$ vs.\ $P$.  Successive approximations without a $P^2$ term in the fitting functions.]{
\resizebox{90mm}{!}{\includegraphics{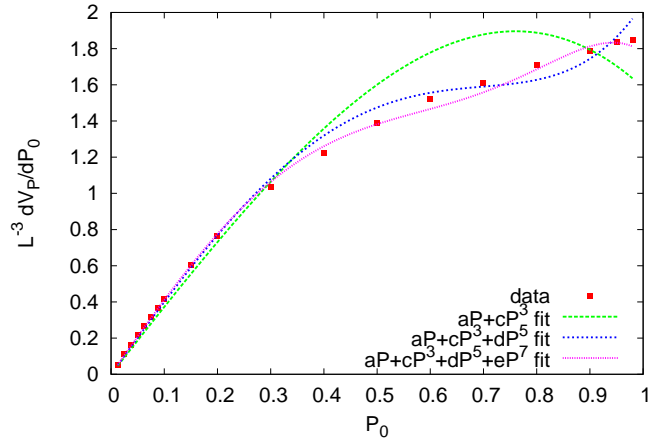}}
\label{sfig4b}
}
\subfigure[~ test if ${dV_P \over dP}$ is an odd function.]{
\resizebox{79mm}{!}{\includegraphics{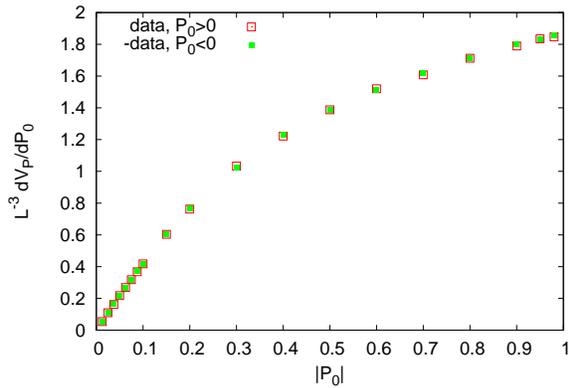}}
\label{sfig4d}
}
\subfigure[~  ${1\over L^3}{dV_P \over d(P^2)}$ vs.\ $P^2$, same data set (and fit) as (a)]{
\resizebox{79mm}{!}{\includegraphics{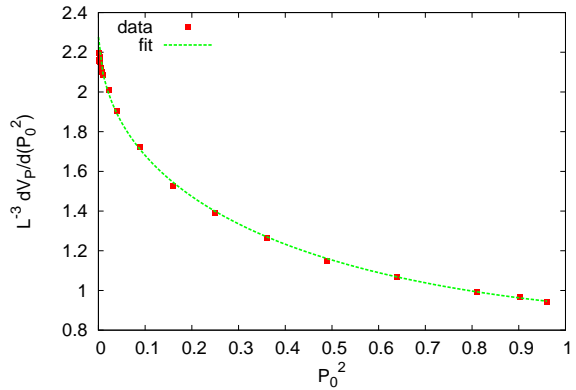}}
\label{sfig4c}
}
\caption{Derivatives of the potential.  Subfigure (a) shows the best fit to the data by a polynomial ${aP + bP^2 + cP^3}$, while 
subfigure (b) shows a best fit by polynomials with two, three, and four odd powers of $P$, which are forms that might be expected from unbroken center symmetry.  (c) is a test of whether $dV_P/dP$ is an odd function of $P$.  Data for the derivative at values of $P_0<0$ are multiplied by -1, for comparison with the data at $P_0>0$. (d) same data (and fit) as in subfigure (a), plotted in a different way.}
\label{fig4}
\end{figure}

      Figure \ref{sfig4a} shows the data for the derivative
\bea
          D(P_0) &\equiv& {1\over L^3} {dV_P \over dP_0} 
\non \\
&=& {1\over L^3} {dS_P \over da} 
\eea
at $\b=2.2$ on a $12^3 \times 4$ volume, which, as in the strong-coupling case, extrapolates linearly to zero at $P_0=0$.   Also
shown is a best fit of $D(P)$ to the polynomial
\bea
   f(P) = c_1 P + c_2 P^2 + c_3 P^3
\label{fitfunc}
\eea
with the best fit constants shown in Table \ref{tab1}.
What is initially a little troubling about this fit is that upon integration, and up to an irrelevant integration constant, we must have
\bea
\V(P_\vx) = \oh c_1 P_\vx^2 + {1\over 3} c_2 P_\vx^3 + \oq c_3 P_\vx^4  \ ,
\eea
which appears to violate center symmetry, i.e.\ $\V(P_\vx) = \V(-P_\vx)$ for SU(2) gauge theory.  Because of center symmetry, the character expansion of $\V(P_\vx)$ contains only characters $\chi_j$ with $j=$ integer.  It is a property
of the SU(2) group characters that each $\chi_j$ can be expressed as a polynomial of order $2j$ in $P$, containing only even powers of $P$ for $j=$ integer, and only odd powers for $j=$ half-integer.   Then if the character expansion of 
$\V(P_\vx)$ is truncated at some $j=j_{max}$, the $P$-derivative is a polynomial in odd powers of $P$ up to 
$P^{2j_{max}-1}$.   

    One might expect that $\V(P_\vx)$ can be accurately approximated by a handful of group characters.
However, the attempt to fit the data with only a few odd powers of $P$ is unsuccessful, in the sense that each of the three fitting functions
\bea
      f(P) = \left\{ \begin{array}{l}
             c_1 P + c_3 P^3 \cr
             c_1 P + c_3 P^3 + c_5 P^5 \cr
             c_1 P + c_3 P^3 + c_5 P^5 + c_7 P^7 \end{array} \right. ~~~ \ ,
\eea
corresponding to truncated character expansions with $j_{max}=2,3,4$, respectively, gives an unacceptable fit, as seen in Fig.\ \ref{sfig4b}.  The reduced $\chi^2$ values in the three cases are  $440,100,25$, respectively.  This is to be compared to the reduced $\chi^2 = 3.2$ for the fitting function \rf{fitfunc}.

\begin{table}[h!]
\begin{center}
\begin{tabular}{|c|c|c|} \hline
\multicolumn{3}{|c|}{Potential fit} \\ \hline
         $c_1 $ &  $c_2$   &  $c_3$   \\ \hline
        4.61(2) &  $-4.51(10)$ & 1.77(8)  \\ \hline
\end{tabular}
\end{center}
\caption{The constants $c_{1-3}$ derived from a best fit of $c_1 P + c_2 P^2 + c_3 P^3$ to the
potential data.}
\label{tab1}
\end{table}

    All this seems to imply that $\V(P_\vx)$ has a term violating center symmetry, but of course that cannot be the case. 
In order that $\V(P_\vx)$ is an even function of $P_\vx$, it must be that the derivative is an odd function, 
$D(P_0)=-D(-P_0)$, which in turn means that the coefficient of the quadratic term in \rf{fitfunc} must change sign when
$P_0 \ra -P_0$. This is easy to check; we simply repeat the calculation with $P_0 < 0$ in \rf{path1}, with the result shown in \ref{sfig4d}. Here the squares are the data for $D(P_0)$ at $P_0>0$, while the circles are data for $(-1)\times D(P_0)$ at $P_0<0$. The fact that the corresponding data points at $\pm P_0$ lie on top of each other means that the derivative is an odd function, and the potential itself is an even function of $P_\vx$, as it must be.  The conclusion, which follows from the best fit, is that over the full range $-1\le P_\vx \le 1$ the potential, up to an irrelevant constant, is given by
\bea
\V(P_\vx) = \oh c_1 P_\vx^2 + {1\over 3}c_2 |P_\vx|^3 + \oq c_3 P_\vx^4  \ .
\label{potential}
\eea
This function is non-analytic, because of the absolute value, but still center symmetric, with the constants given in
Table \ref{tab1}.  It should be emphasized again that this potential cannot be approximated very well by a simple sum of
 $j=0,1,2,3,4$ SU(2) group characters.  Of course, any class function (including $|P_\vx|^3$) can be approximated by a sufficiently large number of group characters, just as a step function can be approximated by a truncated Fourier series.  But keeping only a relatively small number of group characters introduces ``wiggles" in the approximation to the potential (which are seen in Fig.\ \ref{sfig4b}) much like the truncated Fourier series does for the step function.

     So far we have only looked at a pure gauge theory in the confined phase, but it is also possible to compute 
 $\V(P_\vx)$ in the deconfined phase using the same methods.  In comparing the potential in the confining and
 deconfining phases it is useful to display the data in a slightly different way, by plotting the derivative $d V_P/d(P^2)$ vs.\ $P^2$, i.e.
\bea
       {1\over L^3} {dV_P \over d(P_0^2)} = {1\over L^3} {1\over 2P_0}{dV_P \over dP_0}      \ ,
\label{der2}
\eea
When the data is plotted in this way, a curious feature does show up.   First, consider the confined phase.  The data for the above derivative in the confined phase, at the same coupling $\b=2.2$ and lattice volume as before, 
is shown in Fig. \ref{sfig4c}.  In this plot, the best fit shown in Fig.\ \ref{sfig4a} transforms to
\bea
     g(P^2) = \oh (c_1 + c_2 \sqrt{P^2} + c_3  P^2)  \ ,
\label{gfunc}
\eea
with the same constants $c_{1-3}$ shown in Table \ref{tab1}, and this function is also plotted in Fig.\ \ref{sfig4c}.  Note that if the potential didn't have a cubic term, then we would have to omit the term proportional to $\sqrt{P^2}$.
But then the data should fit a straight line in Fig.\ \ref{sfig4c}, which it quite clearly does not.
 
\begin{figure}[t!]
\centerline{\scalebox{0.9}{\includegraphics{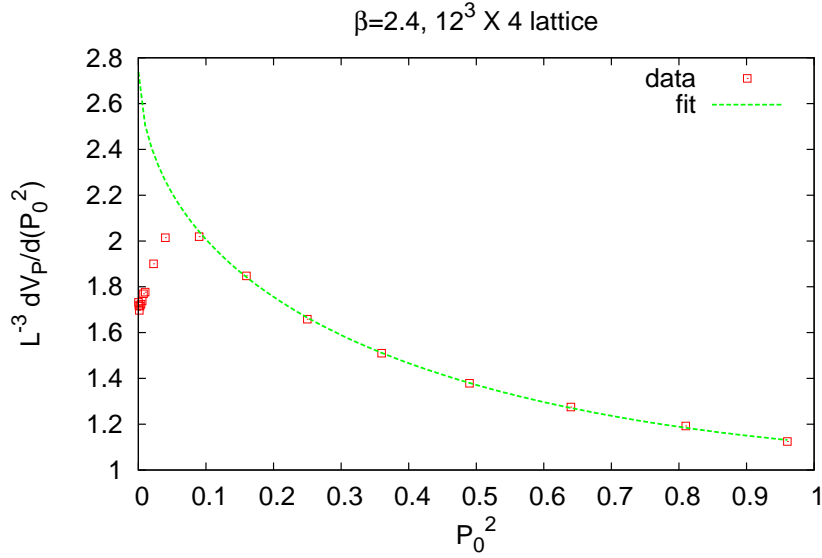}}}
\caption{Derivative of the potential in the deconfined phase.  Note the dip in the data in the
interval ${0<P_0^2<0.1}$. The fit is to data at $P_0^2 \ge 0.1$.}  
\label{fig5}
\end{figure}

   Now we display corresponding data in the deconfined phase. Figure \ref{fig5} shows the result for the derivative 
\rf{der2} at $\b=2.4$, again on a $12^3 \times 4$ lattice, which is well past the deconfinement transition.  Note the peculiar ``dip" near $P_0=0$.  Because of this dip, the polynomial  form 
\rf{fitfunc} to the derivative, which translates to \rf{gfunc} for $d V_P/d(P^2)$, cannot fit the data over the full range.  
It {\it is} consistent with the data away from the dip, i.e.\ at $P_0^2>0.1$, and the resulting fit to data in the interval $[0.1,1]$ is also shown in Fig.\ \ref{fig5}.  The relationship of the dip in the derivative near $P_0=0$ to the deconfinement phenomenon is not obvious to the author.

\begin{figure}[t!]
\centerline{\scalebox{0.9}{\includegraphics{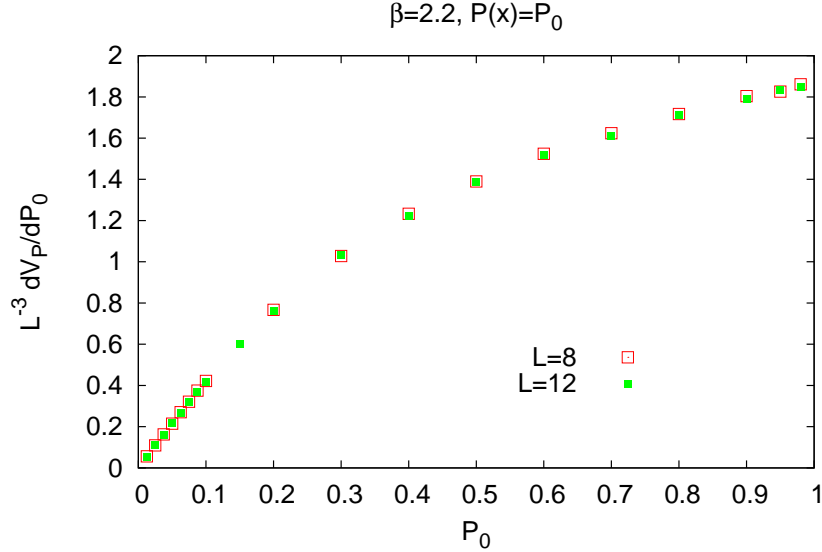}}}
\caption{A test of volume dependence of the potential at $\b=2.2$.  Data for the potential derivative is displayed for lattice
volumes $8^3\times 4$ (open squares) and $12^3 \times 4$ (green circles). }
\label{vol}
\end{figure}

    Finally, it is important to ask whether the potential shown in Fig.\ \ref{fig4} is dependent on the spatial volume.  
In Fig.\ \ref{vol} we show the previous data for the derivative of the potential, obtained on a $12^3 \times 4$ lattice,
together with data for the same observable obtained on an $8^3 \times 4$ lattice.  It can be seen that the volume dependence is negligible in this case.

\section{\label{sec:matter}Potential $\mathbf V_P$ in SU(2) gauge-Higgs theory} 

\begin{figure}[t!]
\centerline{\scalebox{0.7}{\includegraphics{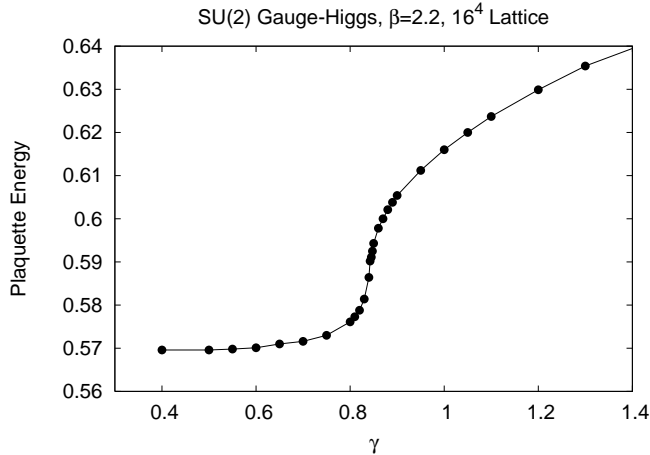}}}
\caption{Plaquette energy vs.\ gauge-Higgs coupling $\gamma$ at fixed $\b=2.2$, for the SU(2) gauge-Higgs theory with fixed Higgs modulus, showing a sharp crossover at $\gamma \approx 0.84$.}  
\label{fig6}
\end{figure}

    We now add a matter field to the gauge theory, to see how this will affect the potential.  To keep the computation requirements very modest, we consider a scalar matter field, in the fundamental representation, with a fixed modulus
(i.e.\ a ``gauge-Higgs" theory).  For the SU(2) gauge group, the matter field can be mapped onto SU(2) group elements, and the action can be expressed as
\bea
    S = \b \sum_{plaq} \oh \mbox{Tr}[UUU^{\dg}U^{\dg}] + \gamma \sum_{x,\m} \oh
              \mbox{Tr}[\phi^\dg(x) U_\m(x) \phi(x+\widehat{\m})]   \ .
\label{ghiggs}
\eea
There have been many numerical studies of this action, following the work of Fradkin and Shenker \cite{Fradkin:1978dv},
itself based on a theorem by Osterwalder and Seiler \cite{Osterwalder:1977pc}, which showed that the Higgs region and
the ``confinement-like" regions of the $\b-\g$ phase diagram are continuously connected.   Subsequent Monte Carlo studies found that there is only a single phase at zero temperature (there might have been a separate Coulomb phase), although there is a line of first-order transitions between the confinement-like and Higgs regions, which eventually turns into a line of sharp crossover around ${\b=2.775,\g=0.705}$, cf.\ \cite{Bonati:2009pf} and references therein.   At $\b=2.2$ the crossover occurs at $\g \approx 0.84$, as seen in the plaquette energy data shown in Fig.\ \ref{fig6}.  There is also a steep rise in the Polyakov line expectation value as $\g$ increases past this point.       

\begin{figure}[htb]
\centering
\subfigure[~ ${1\over L^3}{dV_P \over dP}$ vs.\ $P$ for the gauge-Higgs theory.]{
\resizebox{79mm}{!}{\includegraphics{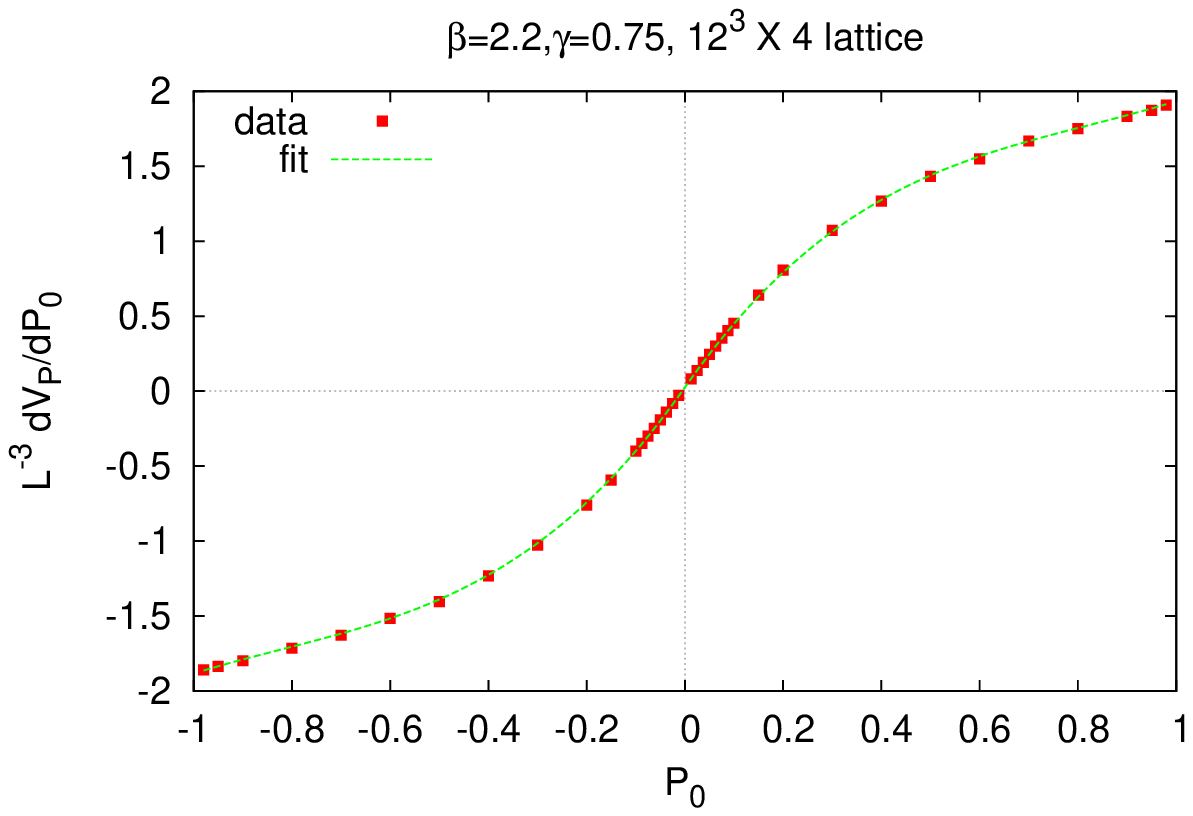}}
\label{fig7a}
}
\subfigure[~ A closeup near $P_0=0$.]{
\resizebox{79mm}{!}{\includegraphics{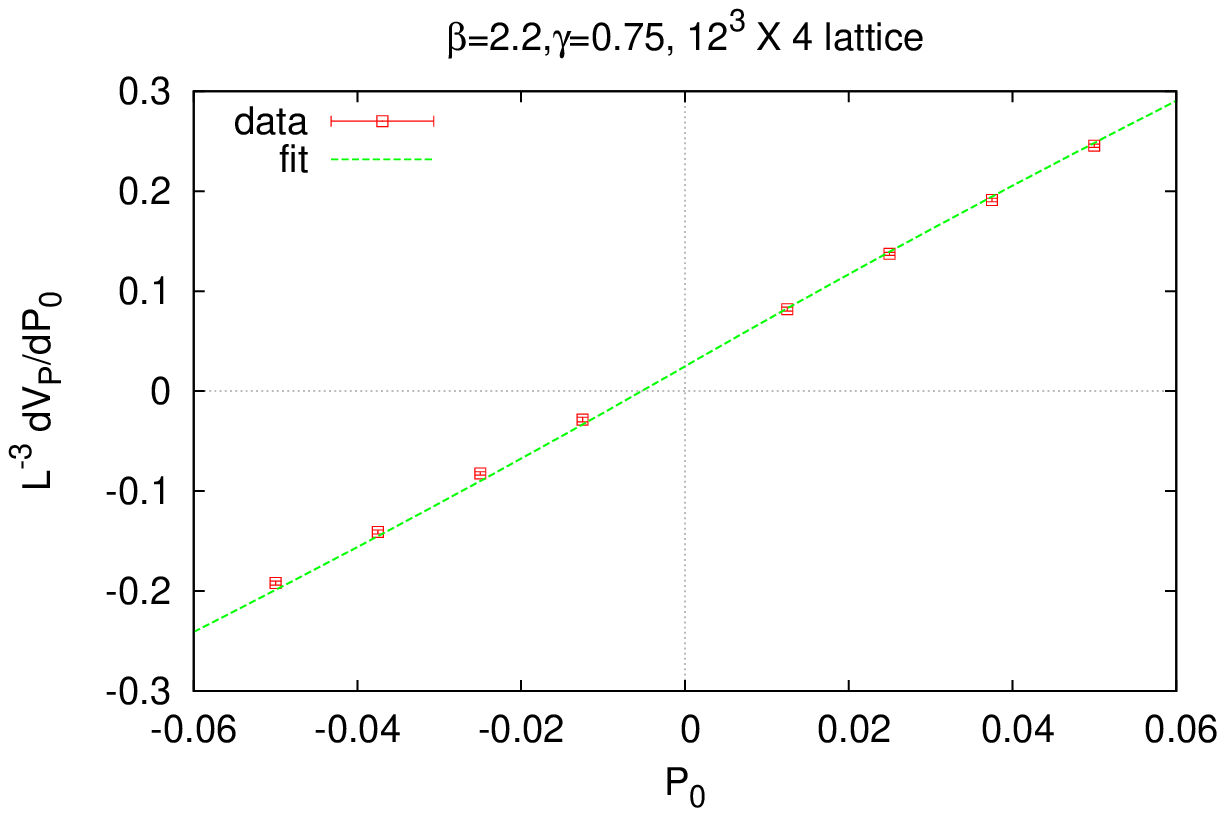}}
\label{fig7b}
}
\caption{Derivative of the Polyakov line potential, per unit volume, with respect to the Polyakov line value $P$, for the 
SU(2) gauge-Higgs theory on a $12^3 \times 4$ lattice.  Data is taken at gauge coupling $\b=2.2$ and gauge-Higgs coupling $\g=0.75$. (a) the data over the range $-1<P<1$, together with the best fit; (b) the data in the vicinity of $P=0$, also showing the fit in this region derived from the full range of data (i.e.\ same curve as in (a)).  Note that the line through the data does not pass through the origin, which implies a small breaking of center symmetry.}
\label{fig7}
\end{figure}

    Fig.\ \ref{fig7a} shows the potential derivative $L^{-3} dV_P/dP_0$ vs $P_0$, along with a best fit to the data,
at $\b=2.2$ and $\g=0.75$,
which is somewhat below the crossover, in the ``confinement-like" regime.  We compute this derivative, again in a
$12^3 \times 4$ lattice volume, at both positive and negative values of $P_0$, to test for the presence of a small
center-symmetry breaking term in the potential (which is not obvious in Fig.\ \ref{fig7a}).  The data over the full range is fit to the form
\bea 
f(P) = c'_0 + c'_1 P + c'_2 \text{sign}(P) P^2 + c'_3 P^3
\eea
which translates, upon integration, into a potential
\bea
\V(P_\vx) = c'_0 P_\vx +  \oh c'_1 P_\vx^2 + {1\over 3}c'_2 |P_\vx|^3 + \oq c'_3 P_\vx^4  \ .
\label{potential1}
\eea
with a center symmetry breaking term $c'_0 P_\vx$.   The constants obtained from the fit are shown in Table \ref{tab1a}.

\begin{table}[h!]
\begin{center}
\begin{tabular}{|c|c|c|c|} \hline
\multicolumn{4}{|c|}{Potential fit: gauge-Higgs model} \\ \hline
     $c'_0$  &   $c'_1 $ &  $c'_2$   &  $c'_3$   \\ \hline
      0.025(1) &  4.70(2) & $-4.70(8)$ &  1.91 (7) \\ \hline
\end{tabular}
\end{center}
\caption{The constants $c'_{0-3}$ derived from a best fit of $ c'_0 + c'_1 P + c'_2 \text{sign}(P) P^2 + c'_3 P^3$ to the
potential data of the SU(2) gauge-Higgs model.}
\label{tab1a}
\end{table}

   The slight asymmetry which breaks $f(P)=-f(-P)$, and therefore center symmetry, is more evident when we 
expand the plot in the immediate region of $P_0=0$, as in Fig.\ \ref{fig7b}.  It can be seen that the best fit through the data points does not go through $f(P_0)=0$ at $P_0=0$, but rather crosses the $y$-axis at a positive value $f(0)=c'_0=0.025$.  The line shown in Fig.\ \ref{fig7b} is taken from a best fit to the full range of data, not just the near $P_0=0$ data.
Since the underlying gauge-Higgs theory breaks center symmetry
explicitly, a term linear in $P_\vx$ is of course expected.  The coefficient $c_0=0.025$ of the symmetry breaking term is quite small, but the expectation value of the Polyakov line at $\g=0.75$ is also quite small:  $\langle P_\vx \rangle = 0.03$ at these couplings and lattice size.   

\section{\label{sec:deform}Plane-wave deformations}

   We now return to the pure gauge theory at $\b=2.2$.  So far the potential term $V_P$ of the effective Polyakov line action has been determined, but the ultimate interest is in the full action.  It was not very hard to extract this action from the $\log[N_n/N_{tot}]$ data at strong couplings.  Unfortunately it is not as easy to jump from the path derivatives to the full action at weaker couplings, simply because $S_P$ is not so simple (and is not known in advance!).   Nevertheless,
knowledge of the action along a particular trajectory in configuration space does provide some information about the
full action.
  
\begin{figure}[ht]
\centering
\subfigure[~ $P_0=0.1$]{
\resizebox{79mm}{!}{\includegraphics{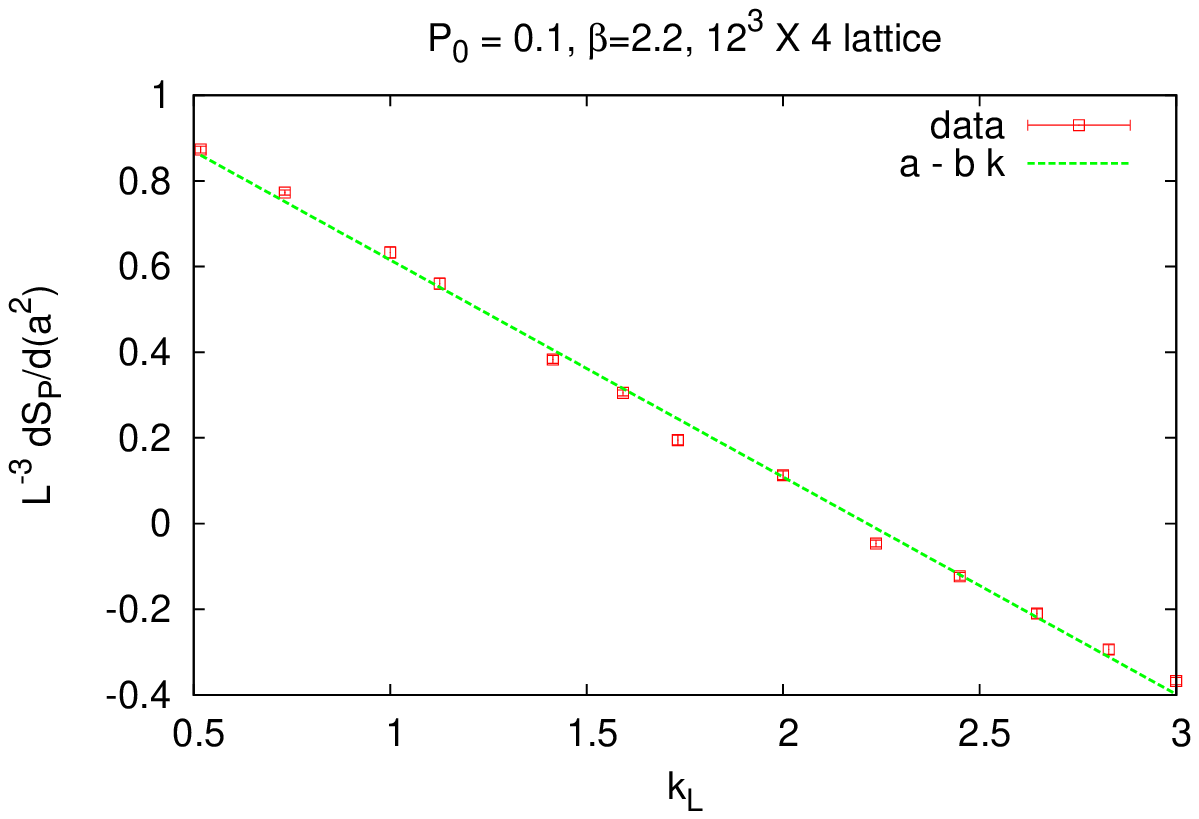}}
\label{add1}
}
\subfigure[~ $P_0=0.8$]{
\resizebox{79mm}{!}{\includegraphics{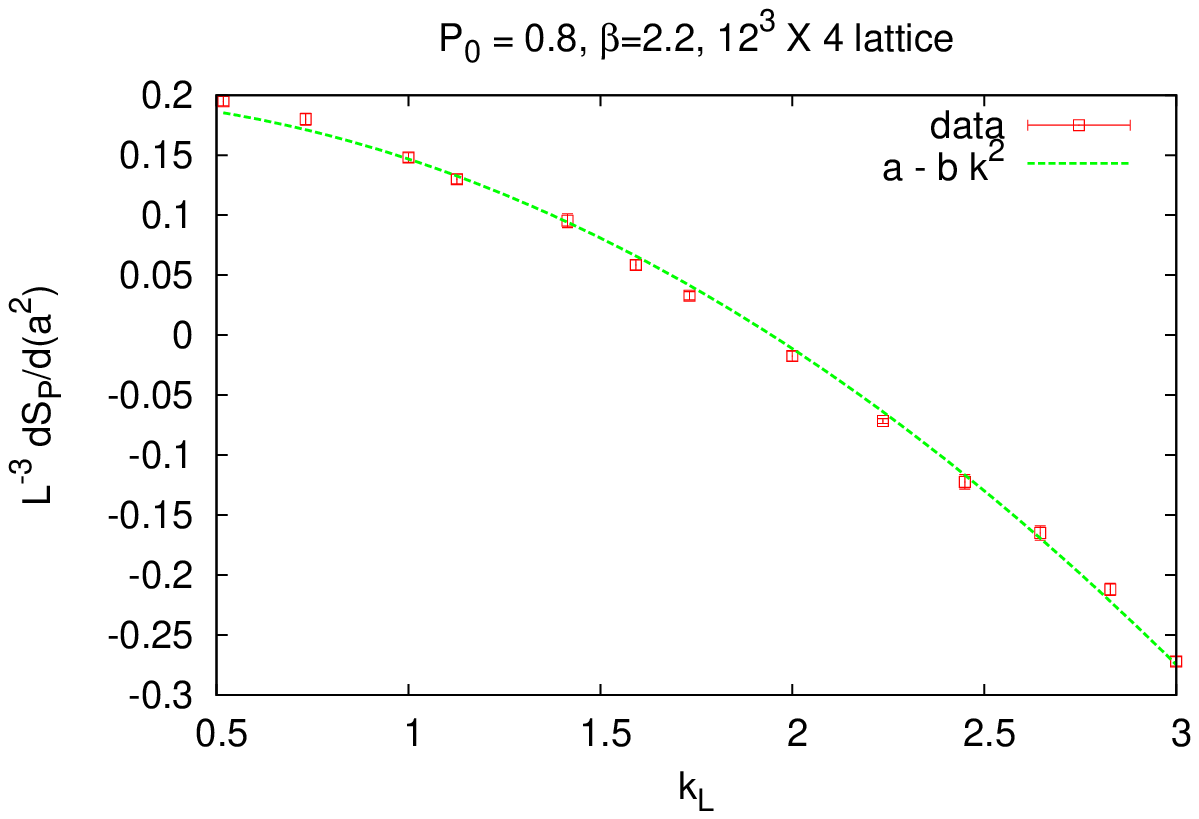}}
\label{add8}
}
\caption{Derivative of the action along a path of plane wave deformations. (a) Data at $P_0=0.1$ is consistent with a linear variation of the derivative with deformation lattice momentum $k_L$; (b) data at $P_0=0.8$ is consistent with a quadratic variation w.r.t.\ $k_L$.}
\label{add}
\end{figure}

\begin{figure}[t!]
\centerline{\scalebox{0.7}{\includegraphics{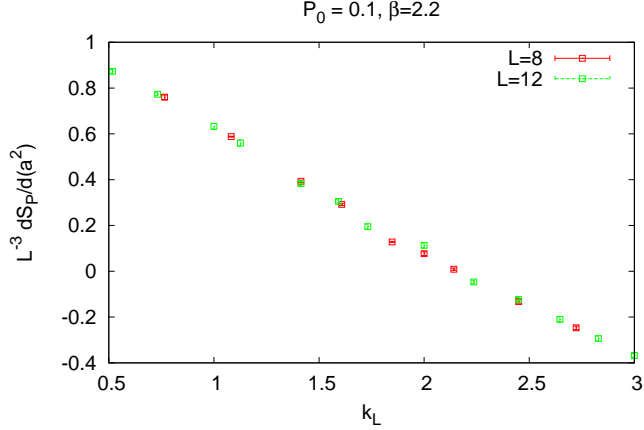}}}
\caption{A check of insensitivity to lattice volume.  Parameters are the same as in Fig.\ \ref{add1}, but this time including data obtained on an $8^3 \times 4$ lattice volume ($L=8$), in addition to data on a $12^3 \times 4$ volume ($L=12$).}  
\label{vol0p1}
\end{figure}

   As in the strong coupling case, we choose to investigate the derivatives of $S_P$ along paths of the form
\rf{trajectory}, i.e.\ plane waves of fixed wavenumber and varying amplitude on a constant background.  The method is the same as outlined in section \ref{sec:test}, but the result is different.  At $\b=1.2$, it was found that
$dS_P/d(a^2)$ was linear in $\kL^2$, and independent of $P_0$.  That is not the case at $\b=2.2$.  What happens in this case is shown in Fig.\ \ref{add}, where we display $L^{-3}dS_P/d(a^2)$ plotted against the magnitude of lattice momentum $\kL=(\kL^2)^{1/2}$ at fixed values of $P_0=0.1$ and $P_0=0.8$.  It can be seen that the $\kL$-dependence of the data in Fig.\ \ref{add1}, at $P_0=0.1$, is consistent with linear, while the $\kL$-dependence in Fig.\ \ref{add8}, at $P_0=0.8$, seems to be quadratic.  This can be seen from fits to $a-bk_L$ in the former case, and to $a-bk_L^2$ in the latter.  This suggests a possible interpolating form
\bea
          {1\over L^3} {d S_P \over d(a^2)}_{|_{a=0}} =   f(P_0) + c \sqrt{\kL^2 + g P_0^2}  \ ,
\label{interp}
\eea
whose $\kL$-dependence would vary continuously from linear, as $P_0 \ra 0$, to quadratic, for ${\kL^2 \ll g P_0^2}$.
Fig.\ \ref{vol0p1} is the same plot as Fig.\ \ref{add1}, except that data obtained on both an $8^3 \times 4$
lattice and a $12^3 \times 4$ volume are displayed together, and both sets of data points appear to have the same
$k_L$ dependence.  This is, of course, evidence of the insensitivity of our results to the spatial volume.

\begin{figure}[t!]
\centering
\subfigure[~ ]{
\resizebox{120mm}{!}{\includegraphics{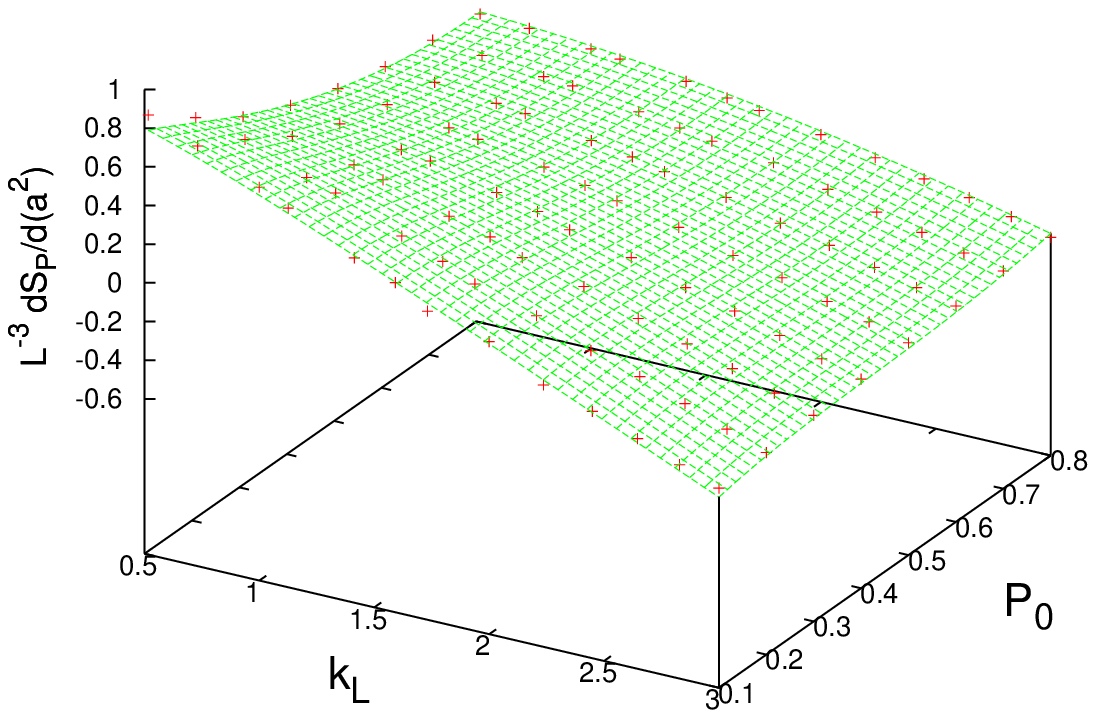}}
\label{kern1}
} 
\subfigure[~]{
\resizebox{120mm}{!}{\includegraphics{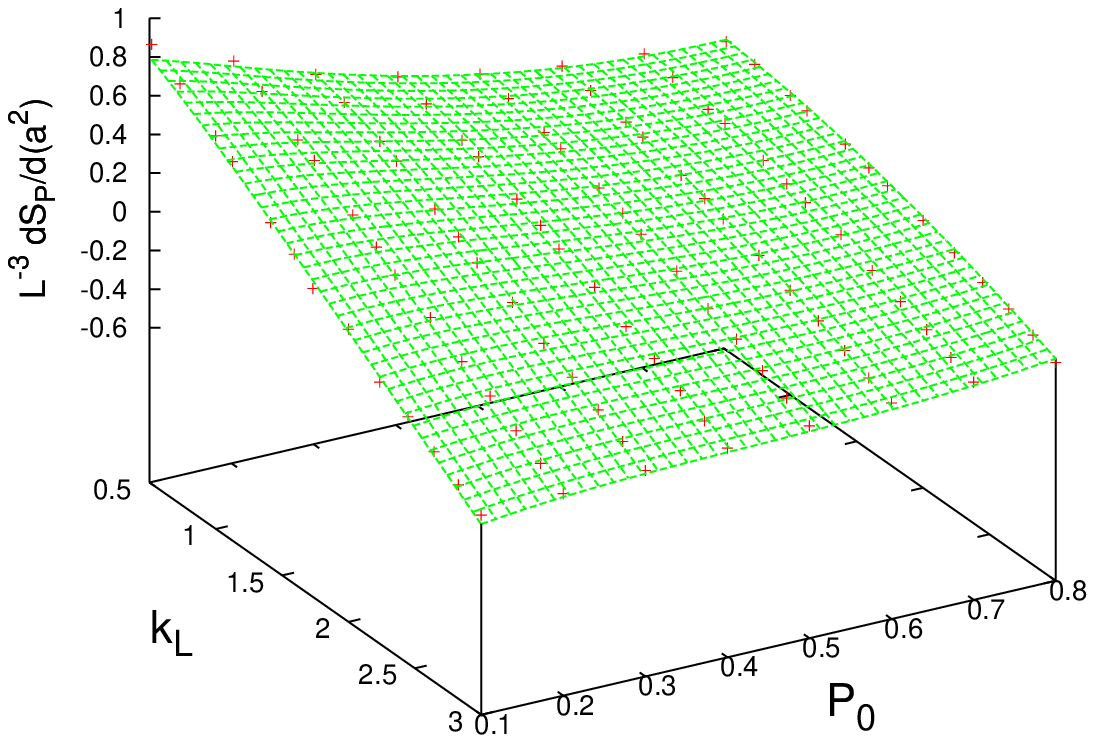}}
\label{kern2}
}
\caption{Two views, at different viewing angles, of the data (red crosses) for $L^{-3}dS/d(a^2)$ vs.\ lattice momentum $k_L$ and Polyakov line $P_0$, and the best fit (green surface) of the form \rf{fitfun} to the data.}
\label{kernel}
\end{figure}

    If \rf{interp} is correct, then it ought to be consistent with the potential \rf{potential}.  This means that $f(P_0)$ can be, at most, quadratic in $P_0$, so let us write
\bea
 {1\over L^3} {d S_P \over d(a^2)}_{|_{a=0}}  = b_0 + b_1 P_0 + b_2 P_0^2 + c \sqrt{\kL^2 + g P_0^2}  \ .
\label{fitfun}
\eea
The constants shown are subject to three constraints by the potential, so if we insist on the potential \rf{potential}
there are really only two independent constants.  In order to derive those constraints, consider a very large lattice volume $L^3$, such that $\kL^2$ can be made very small compared to $g P_0^2$,
but still non-zero, and we assume that $\{c_1,c_2,c_3\}$ do not vary much with $L$ (we have already seen evidence
of this fact in Fig.\ \ref{vol}).  Then the kinetic term is negligible compared to the potential term, and along the trajectory 
\rf{trajectory}, taking account of the spatial average
$(\cos^2 \vk \cdot \vx)_{av} = \oh$,  we have
\bea
      {1\over L^3} {d S_P \over d(a^2)}_{|_{a=0}}  = \oq c_1 + \oh c_2 P_0 + {3\over 4} c_3 P_0^2  \ .
\eea
Comparison with \rf{fitfun} in the $\kL^2 \ll g P_0^2$ limit calls for identifying
\bea
         b_0 = \oq c_1 ~~,~~ b_2 = {3\over 4} c_3 ~~,~~ b_1 + c \sqrt{g} = \oh c_2  \ .
\label{ident}
\eea

    Figure \ref{kernel} show a best fit of the data to the form \rf{fitfun}, with the best fit constants given in Table \ref{tab2}.  This is hardly a perfect fit through the data points, given the value of the reduced $\chi^2 \approx 30$.  Still, except at very low $\kL^2, P_0^2$, the fitting function gives a reasonable account of the dependence of the data on $k_L^2$ and $P_0$.    Table \ref{tab3} is a test of constraints, listing three combinations of constants which, according to the identities \rf{ident}, should vanish.  It is seen that the second and third combinations in the table are consistent with zero, and the first combination is very nearly so.\footnote{All fits, and error estimates on fitting constants, are made using the GNUPLOT software.}

\begin{table}[htb]
\begin{center}
\begin{tabular}{|c|c|c|c|c|} \hline
\multicolumn{5}{|c|}{Surface fit} \\ \hline
          $b_0 $ &  $b_1$   &  $b_2$ & $c$ & $g$    \\ \hline
       1.105(14)  &  0.85(17)  & 1.365(56) & $-0.529(13)$ & 33(3) \\ \hline
\end{tabular}
\end{center}
\caption{Fitting constants $b_{0-2},c,g$ obtained from a best fit to the data points shown in Fig.\ \ref{kernel}, by a surface of the form \rf{fitfun}.}
\label{tab2}
\end{table}

\begin{table}[htb]
\begin{center}
\begin{tabular}{|c|c|c|} \hline
\multicolumn{3}{|c|}{Constraints} \\ \hline 
   $b_0 - \oq c_1$     &  $b_1 + c\sqrt{g} - \oh c_2$   &  $b_2 - {3\over 4} c_3$   \\ \hline
 -0.05(2)    &  0.06(23)  & 0.04(8)   \\ \hline
\end{tabular}
\end{center}
\caption{The constraints \rf{ident} imply that the combination of constants in the second line of the table should vanish within errorbars, and the last line shows the actual values of these combinations, for the constants given in Tables
\ref{tab1} and \ref{tab2}.}
\label{tab3}
\end{table}

\section{\label{sec:towards}Towards the full action}

    The interesting question, of course, is what is the full action which gives rise to the variation \rf{fitfun} along the path, with the given potential \rf{potential}.   We begin by noting that, with the constants shown in Tables \ref{tab1} and
\ref{tab2}, the action
\bea
           S_P &=&  2c \left\{ \sum_{\vx \vy} P_\vx Q_{\vx \vy} P_\vy - \sum_\vx \sqrt{gP_0^2} P_\vx^2 \right\}
           + \sum_\vx \Bigl( \oh c_1 P_\vx^2 + {1\over 3} c_2 |P_\vx^3| + \oq c_3 P_\vx^4 \Bigr)  \ ,
\non \\
          &=& K_P + \sum_\vx \V(P_\vx)
\label{pathS}
\eea
where $K_P$ is the kinetic term
\bea
   K_P = 2c \left\{ \sum_{\vx \vy} P_\vx Q_{\vx \vy} P_\vy - \sum_\vx \sqrt{gP_0^2} P_\vx^2 \right\}
\eea
and
\bea
      Q_{\vx \vy} &=& \Bigl(\sqrt{R}\Bigr)_{\vx \vy}
\non \\
       R_{\vx \vy} &=& (-\nabla_L^2)_{\vx \vy} + gP_0^2 \d_{\vx \vy}
\non \\
                         &=&    \sum_{i=1}^3 (2\d_{\vx \vy} - \d_{\vx,\vy+\ihat} - \d_{\vx+\ihat})  + gP_0^2 \d_{\vx \vy}  \ ,
\label{QR}
\eea 
gives the known results for the potential \rf{potential} and for the variation of $S_P$ with $a^2$ \rf{fitfun} along the paths of plane wave deformations \rf{trajectory}.  The operator $\nabla_L^2$ is the usual lattice Laplacian operator, and $Q$ has the spectral representation
\bea
    Q &=& \sum_\vk \left( \sqrt{k_L^2 + gP_0^2}\right) |\vk \rangle \langle \vk |
\non \\
    Q_{\vx \vy} &=&  {1\over L^3} \sum_\vk \left(\sqrt{k_L^2  + gP_0^2}\right) e^{i\vk \cdot (\vx-\vy)}  \ ,
\label{spectral}
\eea
where $\sum_\vk$ is shorthand for the sum over lattice wave vectors with components $k_i=(2\pi/L) m_i$, and lattice momentum $k_L$ has been defined previously in \rf{k2}.   The ket vectors $|\vk \rangle$ correspond to normalized 
$L^{-3/2} \exp[i \vk \cdot \vx]$ plane wave states.  

   For the paths \rf{trajectory}, set $P_\vx = P_0 + a \cos(\vk \cdot \vx)$, and compute the resulting action on such configurations up to leading order in $a^2$.   Using the spectral representation for the operator $Q$, a short calculation
gives, up to $O(a^2)$,
\bea
    S_P &=&  L^3 \V(P_0) + a^2 L^3  \left\{ \oq c_1 + (\oh c_2 - c\sqrt{g}) P_0 + {3\over 4} c_3 P_0^2 
         + c\sqrt{k_L^2 + gP_0^2} \right\} \ .
\eea
Applying the identities \rf{ident}, which are reasonably well satisfied by the data, this becomes
\bea
    S_P &=& L^3 \V(P_0) 
  + a^2 L^3  \left\{ b_0 + b_1 P_0 + b_2 P_0^2 + c \sqrt{k_L^2 + gP_0^2}   \right\}  \ .
\eea
So we find that for constant configurations ($a=0$), the action is simply the known potential, i.e.\ $S_P=L^3 \V(P_0)$,
while the path derivative is
\bea
{1\over L^3}{dS_P \over d(a^2)}_{|_{a=0}}  = b_0 + b_1 P_1 + b_2 P_0^2 + c\sqrt{k_L^2 + gP_0^2}  \ ,
\label{test}
\eea
in complete agreement with \rf{fitfun}.

    Denote by $P_{av}$ and $\D P^2$ the lattice average value and mean square deviation, respectively, of a given Polyakov line configuration.  It is clear that for the paths \rf{trajectory} considered so far, $P_0=P_{av}$.  One further generalization, which will not affect agreement with the data so far, is to allow the kinetic term to also depend on $\D P^2$, i.e.~\footnote{A generalization of  \rf{pathS} which does {\it not} work is the replacement of $P_0$ by $P_x$ in \rf{pathS} and \rf{QR}.  This leads to additional contributions to $dS_P/d(a^2)$ which spoil the agreement with \rf{test}.}
\bea
   K_P = 2c \left\{ \sum_{\vx \vy} P_\vx \Bigl(\sqrt{-\nabla_L^2 + gP_{av}^2 + g' \D P^2}\Bigr) _{\vx \vy} P_\vy 
          - \sum_\vx \sqrt{ gP_{av}^2 + g' \D P^2} P_\vx^2 \right\}
\eea
It is not hard to see that the O($a^2$) contribution that would arise from the $a^2$-dependence of the square root terms also selects, at this order, the constant $a^2$-independent part of $P_\vx$ and $P_\vy$.  In that case $k_L=0$, and this contribution to the O($a^2$) part of the kinetic term vanishes.  

   In order to investigate the possibility of a $\D P^2$-dependence a little further, let us consider trajectories consisting of plane waves, of varying amplitude $A$, with $P_{av}=0$, i.e.
\bea
              P_x = A \cos(\vk \cdot \vx) \ ,
\eea
and study the derivative $L^{-3} dS_p/dA$ evaluated at $A=A_0$.  To compute this derivative by the relative weights approach,
we construct a set of configurations
\bea
            U^{(n)}_\vx &=& P^{(n)}_\vx \mathbbm{1} + i \sqrt{1 -  (P^{(n)}_\vx)^2} \s_3
\non \\ 
               P^{(n)}_\vx &=&  A_n \cos(\vk \cdot \vx) 
\non \\
 A_n &=& A_0 + \Bigl(n - \oh(M+1)\Bigr) \D A ~~,~~ n=1,2,...,M
\non \\
                   k_i &=& {2 \pi \over L} m_i  \ .
\label{trajectory2}
\eea
and proceed as before.
The conjectured action is 
\bea
S_P &=&    2c \left\{ \sum_{\vx \vy} P_\vx \Bigl(\sqrt{-\nabla_L^2 + gP_{av}^2 + g' \D P^2}\Bigr) _{\vx \vy} P_\vy 
          - \sum_\vx \sqrt{ gP_{av}^2 + g' \D P^2} P_\vx^2 \right\} 
\non \\
     & & + \sum_\vx \Bigl( \oh c_1 P_\vx^2 + {1\over 3} c_2 |P_\vx^3| + \oq c_3 P_\vx^4 \Bigr)
\label{fullS}
\eea 
whose path derivative is~\footnote{The numbers multiplying $c_1,c_2,c_3$ are the lattice averages of $\cos^2(\vk \cdot \vx),
|\cos^3(\vk \cdot \vx)|, \cos^4(\vk \cdot \vx)$ respectively.  These numbers are almost independent of the wavenumber $\vk$ on finite lattices, so long as $\vk \ne 0$, and converge rapidly to the infinite volume limit as lattice volume increases.} 
\bea
{1\over L^3} {dS_P \over dA}_{|_{A=A_0}} &=&  \oh c_1 A_0 + .424 c_2 A_0^2 + .375 c_3 A_0^3 +
   2 c A_0 \left( \sqrt{k_L^2 + \oh g' A_0^2} - \sqrt{\oh g' A_0^2}\right)
\non \\
& &  + \oh c g' A_0^3\left( {1\over \sqrt{k_L^2 + \oh g' A_0^2}} - 
   {1\over \sqrt{\oh g' A_0^2}}\right)
\label{dSa}
\eea

\begin{figure}[t!]
\centerline{\scalebox{0.9}{\includegraphics{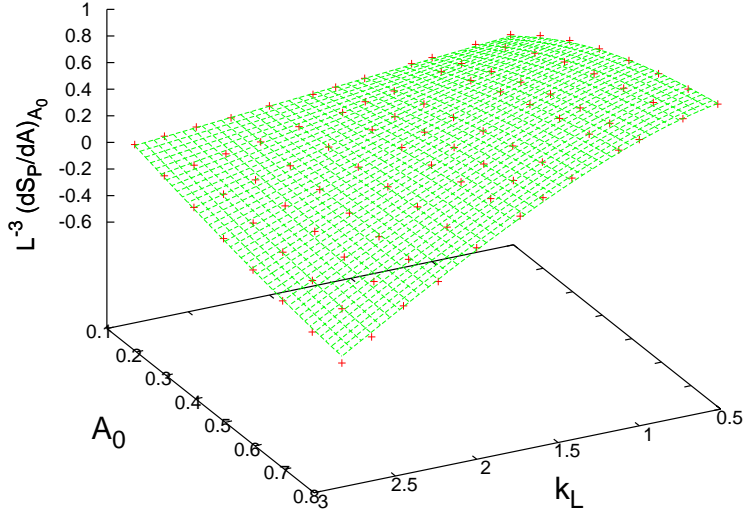}}}
\caption{Variation of Polyakov line action with Polyakov line amplitude, $L^{-3} dS_P/dA$ evaluated at $A=A_0$, for Polyakov line configurations proportional to plane waves $P_\vx = A\cos(\vk \cdot \vx)$, as a function of $A_0$ and lattice momentum
$k_L$.  Red crosses are data points, and the green surface  is a best fit to the data by the analytic form \rf{dSa}.}
\label{pkernel}
\end{figure}

    Taking $c$ and $c_{1-3}$ as given in Tables \ref{tab1} and \ref{tab3}, there is only one free constant left to fit the data, and the best fit, shown in Fig.\ \ref{pkernel}, is obtained at $g'=3.45(4)$.
Once again, this plot should not be interpreted as a perfect fit through the data points within errorbars, given that reduced $\chi^2 \approx 45$.  On the other hand, with only one fitting constant, the expression \rf{dSa} does seem to give a quite reasonable account of the dependence of the data on $A_0$ and $k_L$, despite the highly non-local expression $\D P^2$ introduced into the kinetic term.

\section{Conclusions}

     I have presented a method for computing derivatives $dS_P/d\l$ of the effective Polyakov line action along any given path through field configuration space, parametrized by the variable $\l$.  The technique is easily implemented in a lattice Monte Carlo code by simply replacing updates of timelike links, on a single timeslice, by a Metropolis step which updates that set of links simultaneously, and the potential part $V_P$ of the effective Polyakov line action can be readily   determined, for any given lattice coupling, temperature, and set of matter fields, up to an irrelevant constant.   
It is also possible to determine, from the derivatives, the action $S_P$ along any given trajectory in field configuration space.     
     
     The method has been applied here to SU(2) lattice gauge theory, both without and with a scalar matter field.  At a strong coupling ($\b=1.2$) and finite temperature, the method easily determines the effective Polyakov line action, which we have checked against the known result derived from a strong-coupling expansion.   At a weaker coupling ($\b=2.2$ on a 
$12^3 \times 4$ lattice), where the Polyakov line action is not known, it has been shown that, up to a constant, the potential term has the form
\bea
     V_P =   \sum_\vx \Bigl( \oh c_1 P_\vx^2 +  {1\over 3} c_2 |P_\vx|^3 + \oq c_3 P_\vx^4 \Bigr) \ ,
\eea
with coefficients given in Table \ref{tab1}.  The center-symmetric but non-analytic cubic term comes as a surprise; to the best of my knowledge such a term has not been anticipated in previous studies. It would be interesting to study the evolution of the above potential as $\b$ and $N_t$ vary. Addition of a scalar matter field in the underlying lattice gauge theory introduces a center symmetry breaking term into the potential which is linear in $P_\vx$, with a coefficient reported in section \ref{sec:matter}.  

   Data has also been obtained from small plane-wave deformations around a constant Polyakov line background (Section
\ref{sec:deform}), and for Polyakov lines proportional to a plane waves with variable amplitude (Section 
\ref{sec:towards}).   It was found that the action \rf{fullS} is consistent
with the results that have been found so far, and at this point we may conjecture that \rf{fullS} approximates the desired full
Polyakov line action.  Of course, the kinetic term in $S_P$ could easily have a more complicated form than what is
suggested in \rf{fullS}, and therefore this conjecture needs to be tested on more complicated, non-plane wave configurations.   Those tests, and the extension to the SU(3) group, would be the obvious next steps in the approach introduced here.

\acknowledgments{It is a pleasure to thank Kim Splittorff for many helpful suggestions.  This research is supported in part by the
U.S.\ Department of Energy under Grant No.\ DE-FG03-92ER40711.}

\bibliography{pline}

\end{document}